\documentclass[10pt,twocolumn,letterpaper]{article}

\usepackage{iccv}

\definecolor{darkred}{rgb}{0.7, 0.0, 0.0}

\definecolor{iccvblue}{rgb}{0.21,0.49,0.74}
\usepackage[pagebackref,breaklinks,colorlinks,allcolors=iccvblue]{hyperref}

\def\paperID{7758} 
\def\confName{ICCV}
\def\confYear{2025}

\usepackage{booktabs}
\usepackage{multirow}
\usepackage{graphicx}
\usepackage{caption}
\usepackage{stfloats}

\begin{document}

\title{EvMic: Event-based Non-contact Sound Recovery from Effective Spatial-temporal Modeling}


\author{Hao Yin${^{2,1*}}$
\quad Shi Guo${^{1*}}$
\quad Xu Jia${^{2\dag}}$
\quad Xudong Xu${^{1}}$\\
\quad Lu Zhang${^{2}}$
\quad Si Liu${^{4}}$
\quad Dong Wang${^{2}}$
\quad Huchuan Lu${^{2}}$
\quad Tianfan Xue${^{3,1}}$\\
\normalsize \textsuperscript{1}{Shanghai AI Laboratory} \quad 
\textsuperscript{2}{Dalian University of Technology} \quad 
\textsuperscript{3}{The Chinese University of Hong Kong} \quad \textsuperscript{4}{Beihang University} \\}


\twocolumn[{%
\renewcommand\twocolumn[1][]{#1}%
\maketitle
\begin{center}
    \centering
    \captionsetup{type=figure}
    \includegraphics[width=1.0\textwidth]{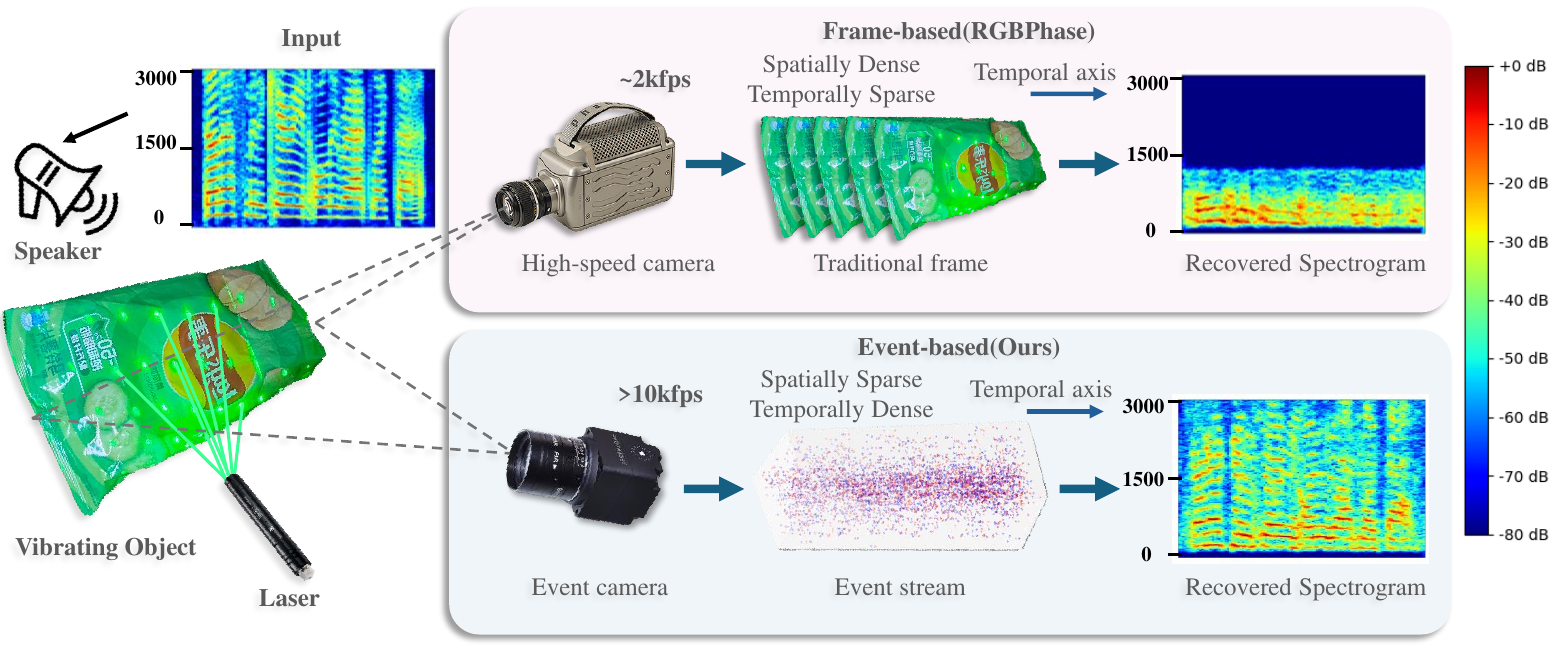}
    \captionof{figure}{Illustration of our event-based non-contact sound recovery. We try to recover sound from the visual vibration of the object caused by the sound wave. Compared with the traditional high-speed camera solution (top), we proposed to use an event camera to capture a temporally dense signal (bottom). We first utilize a laser matrix (left) to amplify the gradient and an event camera to capture the vibrations. Then, our learning-based approach to spatial-temporal modeling enables us to recover better signals.}
    \label{fig:teaser}
\end{center}%
}]
\begin{abstract}
When sound waves hit an object, they induce vibrations that produce high-frequency and subtle visual changes, which can be used for recovering the sound. Early studies always encounter trade-offs related to sampling rate, bandwidth, field of view, and the simplicity of the optical path. Recent advances in event camera hardware show good potential for its application in visual sound recovery, because of its superior ability in capturing high-frequency signals. However, existing event-based vibration recovery methods are still sub-optimal for sound recovery. In this work, we propose a novel pipeline for non-contact sound recovery, fully utilizing spatial-temporal information from the event stream. We first generate a large training set using a novel simulation pipeline. Then we designed a network that leverages the sparsity of events to capture spatial information and uses Mamba to model long-term temporal information. Lastly, we train a spatial aggregation block to aggregate information from different locations to further improve signal quality. To capture event signals caused by sound waves, we also designed an imaging system using a laser matrix to enhance the gradient and collected multiple data sequences for testing. Experimental results on synthetic and real-world data demonstrate the effectiveness of our method. Our project page: \href{https://yyzq1.github.io/EvMic/}{https://yyzq1.github.io/EvMic/}.
\end{abstract}

\renewcommand{\thefootnote}{\fnsymbol{footnote}}
\footnotetext[0]{\hspace{-2em} \textsuperscript{*} indicates equal contributions. This work was done during Hao Yin's internship at Shanghai Artificial Intelligence Laboratory.}
\footnotetext[0]{\hspace{-2em} \textsuperscript{\dag} indicates corresponding author.}

\section{Introduction}
\label{sec:intro}

Non-contact sound recovery has important applications in engineering and scientific fields, such as in surveillance or investigation of the physical properties of materials~\cite{sheinin2022dual, structural2014chen}. This technique mainly relies on cameras to capture subtle deformation of objects caused by sound pressure. As shown in~\cite{davis2014visual}, there is an approximate linear relationship between sound pressure and the displacement of the deformed surface, which shows the feasibility of sound recovery just from visual vibration.

Still, recovering sound signals only from visual cameras remains challenging. This is mainly because object vibrations caused by sound are of both high-frequency and low-amplitude. Vibration frequency often exceeds 1kHz, even faster than some high-speed cameras and the amplitude of vibration is often much less than 1 pixel. Early work~\cite{rothberg1989laser} utilizes laser Doppler velocimeters (LDVs), which have complex optical paths and can only measure a single point. Researchers then explore building a system with simpler optical design~\cite{zizka2011specklesense, zalevsky2009simultaneous, bianchi2014vibration, etchepareborda2021random} by either casting speckle patterns on objects or using phase-based signal processing~\cite{davis2014visual}. Still, these solutions are limited by the bandwidth and sampling rate of frame-based cameras, and are hard to recover sound beyond 1kHz. Recently, sampling rate has been increased by designing different camera systems, either using fast-frame-rate 1D sensors~\cite{wu2020fast,wu202120k,bianchi2019long} or a dual-shutter camera system~\cite{sheinin2022dual}, but these come at the cost of narrower field of view. In short, frame-based cameras have to make trade-offs among three factors: simple optical paths, large field of view, and high sampling rate and bandwidth of the sensors.

Event cameras, characterized by very high temporal resolution, are introduced to achieve a better balance among the above points. Unlike conventional frame-based cameras that capture all pixels' intensity at a fixed frame rate, event cameras only record pixels with certain levels of brightness changes. Because of this sparsity, event cameras can both cover a large field of view and capture temporally dense signals. These properties make event cameras appealing for high-frequency vibration analysis.

To utilize such characteristics of event signals for vibration analysis, \citet{dorn2018efficient} and \citet{niwa2023live} adapted the phase-based method to estimate vibration from the event stream, but they are limited to vibrations along a single orientation. Zero crossing detection is adopted in \cite{howard2023event} for event-based sound recovery but is quite sensitive to noise. These solutions did not fully exploit spatial-temporal information from event streams and did not use deep learning to utilize data prior.

In this work, we propose a novel non-contact sound recovery system based on event cameras. Unlike previous event-based vibration analysis, we propose a novel network structure that can jointly model long temporal dependencies and spatial relationships among events, resulting in high-quality sound recovery. To leverage the power of data-driven techniques, we also collect a new dataset for event-based sound recovery. Since collecting paired data for the non-contact sound recovery task in the real world is challenging, we resort to synthetic data. Specifically, we introduce \emph{EvMic}, the first synthetic dataset for event-based non-contact sound recovery, created using Blender~\cite{blender} and an event simulator~\cite{hu2021v2e}. Furthermore, to validate the effectiveness of the proposed method, we also collected event data in real-world scenarios for evaluation. Since the vibrations on the objects hit by the high frequency sound wave are extremely subtle, we design an imaging system that utilizes a laser matrix to amplify the object’s surface gradients thereby capturing more informative event signals. The laser matrix, comprising multiple laser points, also preserves the advantage of a large field of view. To the best of our knowledge, this is the first work that investigates the potential of event cameras for sound recovery using deep learning techniques.

The proposed network consists of three parts. First, sparse convolution is adopted to leverage the sparsity nature of event data for lightweight visual feature extraction. Second, we use the Mamba model to extract temporal information from the event stream, utilizing its advantage in modeling long sequences. Third, we propose a speckle-based spatial aggregation block to further improve robustness. When sound waves interact with the object's surface, they produce spatially varying vibrations. The magnitudes and directions of vibrations, when projected onto the image plane, are modulated by the local surface normal orientations, and we utilize this property to build the spatial aggregation block.



\section{Related work}
\textbf{Non-contact sound recovery}. The non-contact sound recovery aims to measure object vibrations in a non-contact manner using vision sensors. Early studies~\cite{rothberg1989laser} utilize laser Doppler velocimeters(LDVs) which offer high precision but are limited to measuring vibrations at a single point and require a complex optical path. The roughness of the object's surface can create speckle patterns with the laser. Some studies~\cite{zizka2011specklesense, zalevsky2009simultaneous, bianchi2014vibration, etchepareborda2021random} utilize frame-based high-speed cameras to capture speckle patterns to address the shortcomings of LDVs. However, reliance on high-speed cameras introduces limitations in sampling rate and bandwidth. Some works attempt to use 1D sensors. They achieved a higher sampling rate but at the expense of the field of view.  Additionally, they can only measure vibrations in a single direction. Sheinin \textit{et al.}~\cite{sheinin2022dual} introduced a dual shutter imaging system, replacing the high-speed camera with two simultaneously low-speed cameras equipped with rolling and global shutter sensors. This approach no longer requires high-speed cameras but the arrangement of laser points is restricted by the dual shutter system.  Davis \textit{et al.}~\cite{davis2014visual} directly captured vibrating objects with high-speed frame-based cameras and designed a phase-based method to recover sound waves. This method eliminated the optical path of LDV and achieved a larger field of view but is still limited by high-speed frame-based cameras. Our work captures vibrating objects using event cameras. This allows us to simultaneously obtain high-frequency data and a large field of view while maintaining a simple optical path.\\
\textbf{Event-based non-contact vibration measurement.} Event cameras trigger each pixel asynchronously, allowing for extremely high temporal resolution and eliminating motion blur. Additionally, event cameras detect only areas where brightness changes occur, thereby avoiding the capture of static background information and significantly reducing bandwidth usage. These features make event cameras exceptionally well-suited for vibration measurement. Some works attempt to detect micro vibrations~\cite{pfrommer2022frequency, na2023event, 2024EVmotionmag, lv2024structural, bane2024non}. Ge \textit{et al.}~\cite{ge2022lens, ge2020dynamic} employed speckle patterns to analyze micro displacements. Shi \textit{et al.}~\cite{Shi2023EvVibrationLaser} used Laser-Assisted Illumination and proposed a method based on a mixed Gaussian distribution to analyze signals of up to 300 Hz. Dorn \textit{et al.}~\cite{dorn2018efficient} adopted a phase-based method with event streams for vibration analysis, achieving high precision but not attempting to recover high-frequency signals. \citet{niwa2023live} and \citet{howard2023event} further explored the use of an event camera for non-contact sound recovery. 
These methods~\cite{niwa2023live,dorn2018efficient,howard2023event}, offer a broad field of view but do not fully utilize the spatial-temporal information. In this work, we propose a learning-based approach for effective spatial-temporal modeling, thereby enabling the recovery of clearer sound waves.\\

\section{Problem formulation}
\textbf{Event representation}. 
Event cameras are bio-inspired sensors that asynchronously capture changes in brightness at each pixel location. An event is triggered when the change in log brightness exceeds a predefined threshold $C$, and its polarity $p$ is defined as:
\begin{equation}
    p = \begin{cases}
        1, & \text{if } L\left (u,t\right ) - L\left (u,t - \Delta t\right) \geq C, \\ 
        -1, & \text{if } L\left (u,t\right ) - L\left (u,t - \Delta t\right) \leq -C, \\        
        0, & \text{otherwise},
        \end{cases}       
\label{eq:normal_define_evs}
\end{equation}
where $L$ represents the log brightness, $u = \left( x, y \right)$ denotes the pixel coordinates, $t$ is the timestamp. The output from the event camera over a given time period is a set of events, defined as $\left \{ e_{k} \right \}  = \left \{ u_{k}, t_{k}, p_{k} \right \} _{k=1}^{K} $. K denotes the total number of events.

\noindent \textbf{Event-based non-contact sound recovery}. 
When sound waves strike an object, the induced sound pressure causes slight deformations on its surface, resulting in vibrations that synchronize with the frequency of the sound wave. These vibrations lead to variations in image brightness, which in turn trigger high-frequency event signals. The primary objective of this task is to analyze the event signals generated by object vibrations and to recover the sound information from these visual signals.

To formalize this process, we first define the pressure of the sound wave at time $t$ as $q_{t}$. \citet{davis2014visual} demonstrated that the displacement of an object vibrating under a sound wave of a certain frequency is approximately linearly related to the sound pressure as: \begin{equation} 
\delta_{t-1\to t}=\alpha (q_{t} - q_{t-1}), \end{equation} 
where $\delta_{t-1\to t}$ represents the motion field over the time interval between $t-1$ and $t$, and $\alpha$ is the linear coefficient that depends on the material properties and the frequency of the sound wave. Under the assumption of constant illumination, for small $\Delta t$, the logarithmic change in brightness can be approximated as~\cite{gallego2020event}:
\begin{equation} 
\Delta L \approx - L' \cdot \delta_{t-1\to t} \approx - L' \cdot (\alpha (q_{t} - q_{t-1})), 
\label{eq:bright_change_with_vibrant} 
\end{equation} 
where $L'$ denotes the gradient of $L$. By substituting the event representation from \cref{eq:normal_define_evs} into \cref{eq:bright_change_with_vibrant}, we obtain the relationship between sound pressure and event signal, expressed as: \begin{equation} 
p_k C \approx - L' \cdot (\alpha (q_{t} - q_{t-1})). 
\label{eq:event_with_sound_pressure} 
\end{equation}

The goal is to recover the sound pressure $q_{t}$ from the event stream $\{p_k\}$, thereby reconstructing the sound wave. High-frequency vibrations typically correspond to smaller sound pressure changes $\Delta q = q_{t} - q_{t-1}$. Frame-based cameras record absolute intensity to capture subtle changes, while events are generated by the combined effects of gradients and subtle sound pressure changes (see \cref{eq:event_with_sound_pressure}). To enhance the event signal, we use a laser to enhance the object's surface gradient  $L'$ (as illustrated in \cref{fig:teaser}).

\begin{figure}[htbp]
    \centering
    \includegraphics[width=0.49\textwidth]{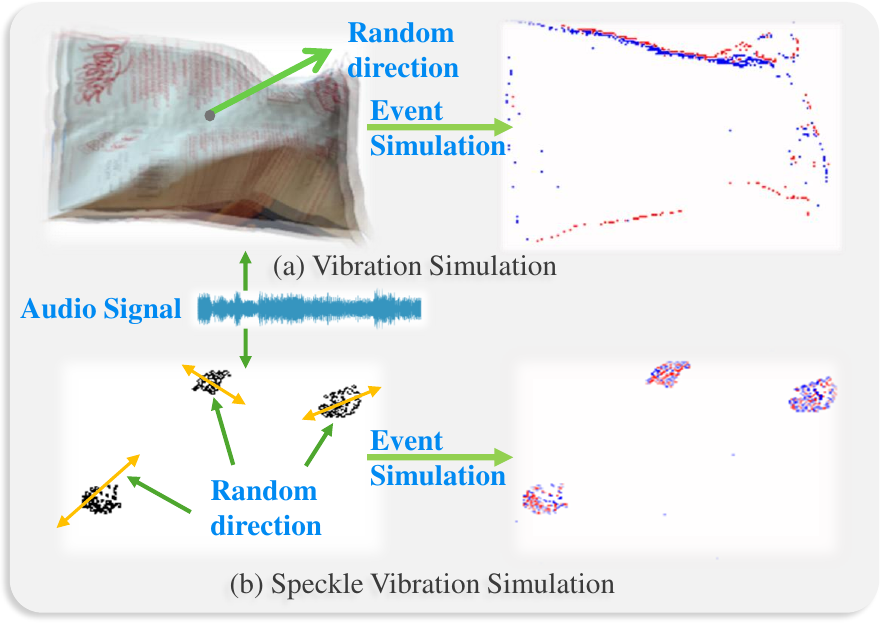}
    \caption{(a) Our data simulation starts with controlling the objects' vibration.  We utilize audio to manipulate the coordinates of objects resulting in their vibrations across random directions. Then we use an event simulator to generate the corresponding events. The generated events are used for training. (b) The synthetic vibrating speckles are used for fine-tuning and testing.}
    \label{fig2}
\end{figure}

\section{Method}

\begin{figure*}[!ht]
    \centering
    \includegraphics[width=1.0\textwidth]{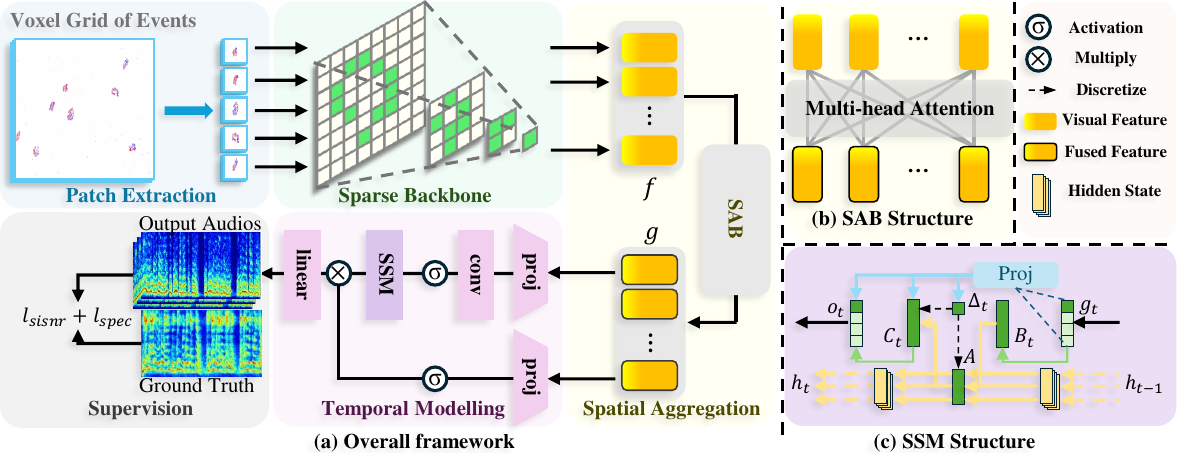}
    \caption{(a) Overview of our proposed network architecture. The event stream is processed into voxel grids, from which patches centered around the speckles are selected. First, the patches are input into a sparse convolution-based lightweight backbone to extract visual features. Next, a spatial attention block aggregates the information in the different patches. Finally, Mamba is employed to model long-term temporal information and reconstruct the audio that caused the object’s vibration. (b) and (c) illustrate the detailed structure of SAB and SSM. (c) At time t \(g_t\) is the input feature, \(o_t\) is the output and \(h_t\) denotes the hidden state. A, B, and C are the gating weights optimized by Mamba. $\Delta$ is used to discretize the continuous parameters $A$ and $B$.}
    \label{fig3}
\end{figure*}

\subsection{Data simulation}
\label{sec:Dataset}
In real-world scenarios, obtaining ground truth for non-contact sound recovery tasks is challenging due to alignment issues, reverberation, and the object's unknown response to sound. Inspired by other tasks~\cite{oh2018learning,2024EVmotionmag}, we create EvMic, a simulated dataset that contains approximately 10,000 segments of data.

The simulation pipeline begins with synthesizing high-frame-rate video sequences. We use Blender~\cite{blender} to render realistic visual effects. The key aspect of the simulation is controlling the object's vibrations according to the audio, as shown in Fig.~\ref{fig2} (a). Specifically, we generate a random direction in Blender, with the audio signal serving as the vibrational displacement along it. Moreover, the audio will be multiplied by a random gain. The rendered video (346×260 resolution) exhibits both pixel- and sub-pixel-level vibrations based on audio and gain. We then simulate events using V2E~\cite{hu2021v2e}, with the results visualized in Fig.~\ref{fig2}(a).

To further bridge the gap between sim and real, a set of supplementary data is synthesized in the same manner as shown in Fig.~\ref{fig2} (b). Multiple speckles are controlled to vibrate according to the same audio signal, with each exhibiting unique vibration directions and signal gains.  These additional data are utilized to finetune the spatial aggregation block, enhancing the model's generalization ability.

\subsection{Network architecture}
\label{sec:network}
Our network is designed to estimate sound from event signals that exhibit temporally dense yet spatially sparse characteristics, as illustrated in Fig.~\ref{fig3}. Events are converted into spatio-temporal voxel grids for neural network processing, following the method in \cite{zihao2018unsupervised}, a common usage in event-based vision. Due to the high frequency of the sound signals, a large number of bins in the event voxel grids are required to retain fine-grained temporal information. Since events triggered by sound vibrations are primarily concentrated around laser speckles, patches centered on these speckles are first extracted using contour extraction. These patches are denoted as $P=\left \{  P_{i} \right \} _{i=1} ^{N}$, where $N$ is the speckle number. Each patch $P_{i}$ has the dimensions $T\times 2\times pw_i\times ph_i$, where $pw_i\times ph_i$ is the patch size, $2$ represents distinct voxel grids containing $e^{+}$ and $e^{-}$ for positive and negative events respectively. $T$ is the length of the time series, equal to the number of bins in the event voxel grid, which is set to 4K in our training process.

To estimate sound from the speckle patches $P=\left \{  P_{i} \right \} _{i=1} ^{N}$, our network comprises three main components: a feature extraction module (extracting features from each patch individually), a spatial aggregation module (aggregating information across patches), and a temporal modeling module (capturing long-term temporal information). Leveraging the sparsity of event data, a sparse convolutional backbone is implemented for efficient visual feature extraction. In the spatial aggregation module, a multi-head self-attention strategy is designed to adaptively aggregate spatial information across different vibration directions. Finally, to efficiently capture long-term temporal dependencies, we introduce a mamba-based~\cite{gu2023mamba} temporal modeling module. Details of these three key components are provided below.




\noindent\textbf{Lightweight visual feature extraction.}
Since each patch $P_i$ has a large temporal dimension, exceeding 2K, visual feature extraction introduces significant computational overhead. Fortunately, event data is inherently sparse, and the small amplitudes characteristic of audio-induced vibrations further contribute to this sparsity. To effectively harness this sparsity and reduce computational costs, we developed a ResNet18~\cite{he2016deep} network based on sparse convolutional layers (SPconv)~\cite{liu2015sparse, graham2017submanifold, zhang2025evsign} to extract visual features, denoted as $f=\left \{  f_{i}\right \} _{i=1} ^{N} \in  \mathbb{R}^{N\times T\times C} $, where $C$ is the feature dimension. Sparse convolutions perform computations only in non-zero regions, which means calculations occur solely at the locations where events are generated, significantly alleviating the computational burden on the model.

\noindent\textbf{Spatial aggregation block (SAB).} To robustly estimate sound vibration, it is important to spatially aggregate features extracted at each local region to increase the signal-to-noise ratio. However, the varying normal directions on the object's surface cause vibrations along different directions with varying amplitudes in the image plane. Using the entire voxel for aggregation averages patches across these directions, potentially leading to signal interference.

To address this issue, we propose an adaptive spatial aggregation module utilizing multi-head self-attention (MSA)~\cite{vaswani2017attention}. For each timestamp $t$ of the input feature $f$, the feature is represented as $f_t \in \mathbb{R}^{N \times C}$. First, projections $Q_t = F_q(f_t)$, $K_t = F_k(f_t)$, and $V_t = F_v(f_t)$ are computed using the projection matrices $F_q$, $F_k$, and $F_v$. The aggregated feature $g_t$ is then calculated as:
\begin{equation}
    g_t = \text{SoftMax}\left(\frac{Q_t K_t^T}{\sqrt{d}}\right) V_t,
\end{equation}
where $d$ is the normalization parameter. Finally, we obtain the aggregated feature across all patches and timestamps, denoted as $g \in \mathbb{R}^{N \times T \times C}$. To train the spatial aggregation module (SAB), we use the simulated vibrating speckles illustrated in Fig.~\ref{fig2} (b).\\
\textbf{Temporal information modeling.} 
Audio sequences exhibit long-term temporal correlations, which are challenging to model. Inspired by the success of structured state space models (SSMs)~\cite{gu2021efficiently, gu2023mamba} in capturing long-term temporal information, we develop a Mamba-based~\cite{gu2023mamba} temporal aggregation module to capture the temporal dependencies in the features and generate the corresponding audio signal. The detailed structure is shown in Fig.~\ref{fig3} (c).

For the input feature of all patches at each timestamp, denoted as $g_i \in \mathbb{R}^{N \times C}$, the output can be modeled as:
\begin{equation}
h_t = \bar{A} h_{t-1} + \bar{B} g_t, \quad o_t = C h_t,
\end{equation}
where $h_t$ is the hidden state at time $t$, $o_i$ denotes the output audio signal from the $i$-th patch, 
\begin{equation}
    \bar{A} = \exp(\Delta A), \quad \bar{B} = (\Delta A)^{-1} (\exp(\Delta A) - I) \cdot \Delta B.
\end{equation}
Here, $A \in \mathbb{R}^{N \times N}$, $B \in \mathbb{R}^{N \times 1}$, and $C \in \mathbb{R}^{1 \times N}$ are learnable parameters, and $\Delta$ is used to discretize the continuous parameters $A$ and $B$.

\subsection{Loss function} 
\textbf{Scale-invariant source-to-noise ratio (SISNR) loss.} The SISNR loss~\cite{luo2019conv} is used to assess the similarity between two signals. Compared to traditional SNR loss, SISNR loss is less affected by variations in volume, defined as,\\
\begin{equation}
L_{\text{sisnr}} = - 10 \log_{10} \left( \left\| \alpha \cdot \hat{o} \right\|_{2}^2 / \left\| \hat{o} - o \right\|_{2}^2 \right),
\end{equation}
where $o$ and $\hat{o}$ are the predicted and ground-truth audio signal, and $\alpha =\langle \hat{o}, o \rangle / \| o \|_{2}^2$.

\noindent\textbf{Multi-scale spectral reconstruction loss.} 
To supervise audio signals in the frequency domain, we define a multi-scale spectral reconstruction loss to enhance the reconstruction of high-frequency components in the audio signal~\cite{soundstream}: 
\begin{multline}
    L_{\text{spec}} = \sum_{s=2^6,2^7\cdots2^{9}}  \left \| S^{s}\left ( o \right ) , S^{s}\left ( \hat{o} \right )  \right \|_{1}  \\+ \alpha_s \cdot \left \| log (S^{s}\left ( o \right )) , log( S^{s}\left ( \hat{o} \right )  \right) \|_{2},  
\end{multline}
where $S^{s}(o)$ denotes the Mel-spectrogram of the audio signal $o$ with scale $s$. In our setting, $s$ is adjusted to \(s=2^{6},2^{7} \cdots 2^{9}\) according to the sampling rate. Additionally, \(\alpha _{s} =\sqrt{s/2} \) serves as a scale-dependent weighting factor. 

The total loss is defined as $L_{\text{total}} = L_{\text{sisnr}} + \beta L_{\text{spec}},$,
where \(\beta = 1e-4\) is set in our experiments. 

\section{Experiments}
\label{sec:Experiments}

\subsection{Training details}
Experiments are conducted using the PyTorch framework~\cite{paszke2019pytorch} on a single NVIDIA 4090 GPU, training the model on the proposed EvMic dataset. The training process is divided into two phases to ensure stable training: First, pre-training the modules without the spatial aggregation block (SAB). The entire network is then fine-tuned. During the training process, predefined regions are used to partition the patches. SGD~\cite{bottou2010large} was used as the optimizer with a learning rate of $10^{-4}$. A 0.5-second event stream is divided into 2000 bins for input, which corresponds to a temporal resolution of 4000 Hz. The batch size is set to 1, with 200k iterations for pre-training and another 200k iterations for fine-tuning.

\subsection{Baseline methods}
\label{sec:Comparation with other methods}
Since there are currently no public datasets or code for event-based sound recovery, we implemented the state-of-the-art method by \citet{dorn2018efficient} as our baseline (referred to as EvPhase) and also included the frame-based approach by \citet{davis2014visual} (referred to as RGBPhase). To ensure a fair comparison, we fixed the filter parameters of EvPhase, setting \(n=13,\sigma=3,\lambda=32,\gamma=1,\phi=0,\theta=\frac{\pi}{2} \), based on the description in \cite{niwa2023live}.


Following \cite{davis2014visual}, we also applied denoising methods as post-processing for the audio outputs of all comparison methods. For speech data, we used FullSubNet-plus~\cite{chen2022fullsubnet+}, while spectral subtraction~\cite{boll1979suppression} was employed for other types of data. A Butterworth high-pass filter was then applied to remove low-frequency noise.

\textbf{All audio results are provided in supplementary} and we encourage readers to listen to them.

\subsubsection{Results on the synthetic data}

\begin{table*}[!ht]
    \centering
    \caption{Quantity comparison results of our model with other methods on the synthetic data. }
\begin{tabular}{lllllcclcclcclcccc}
\toprule
 & \multicolumn{4}{l}{\multirow{2}{*}{Methods}} & \multicolumn{2}{c}{Female-SA1} &  & \multicolumn{2}{c}{Male-SA1} &  & \multicolumn{2}{c}{Female-SA2} &  & \multicolumn{2}{c}{Male-SA2} & \multicolumn{2}{c}{Average} \\ \cline{6-18} 
 & \multicolumn{4}{l}{} & SNR↑ & STOI↑ &  & SNR↑ & STOI↑ &  & SNR↑ & STOI↑ &  & SNR↑ & STOI↑ & SNR↑ & STOI↑ \\ \hline
 & \multicolumn{4}{l}{RGBPhase} & -2.992 & 0.389 &  & -2.976 & 0.419 &  & -2.578 & 0.246 &  & -2.801 & 0.237 & -2.837 & 0.322 \\
 & \multicolumn{4}{l}{EvPhase} & -1.883 & 0.286 &  & 0.183 & 0.485 &  & 1.465 & \textbf{0.482} &  & -0.080 & 0.251 & -0.079 & 0.376 \\ \hline
 & \multicolumn{4}{l}{Ours(8kHz)} & \textbf{1.159} & \textbf{0.451} &  & \textbf{0.809} & \textbf{0.542} &  & \textbf{1.929} & 0.479 &  & \textbf{0.959} & \textbf{0.452} & \textbf{1.214} & \textbf{0.481} \\ \bottomrule
\end{tabular}

\label{tab1}
\end{table*}

\begin{figure*}[htbp]
    \centering
    \includegraphics[width=1.0\textwidth]{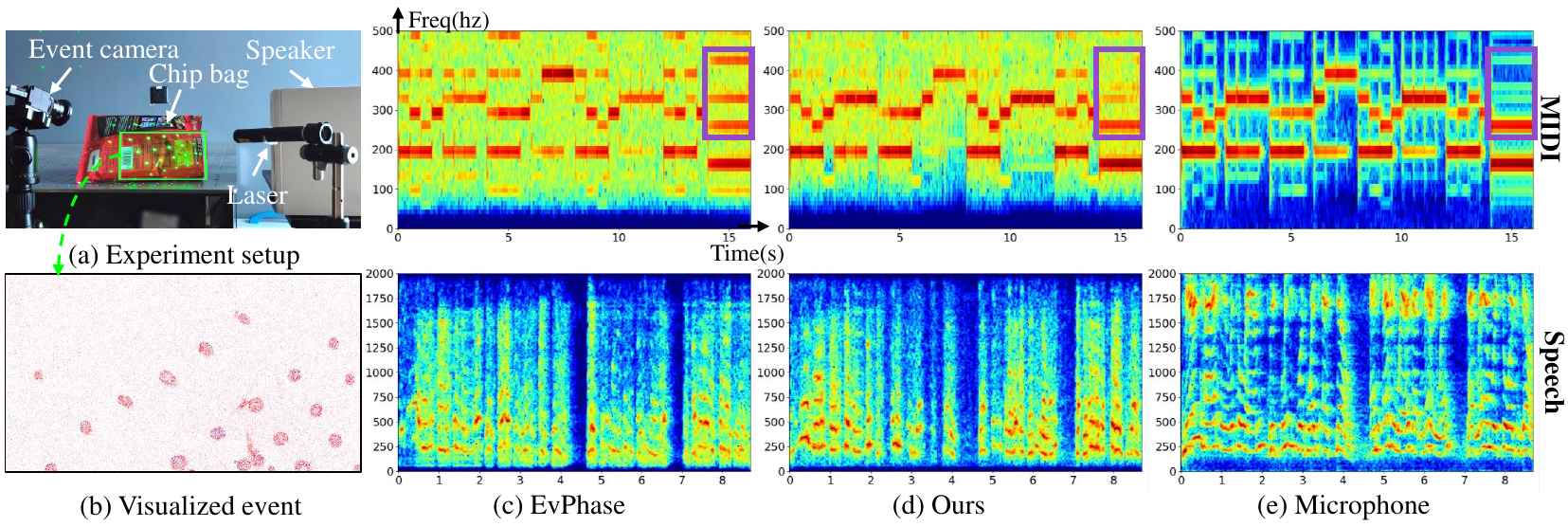}
    \caption{Qualitative comparison results on the real-world data of a chipbag. \textbf{Audio is provided in the supplementary.}}
    \label{fig_chipbag}
\end{figure*}

\begin{figure*}[htbp]
    \centering
    \includegraphics[width=1.0\textwidth]{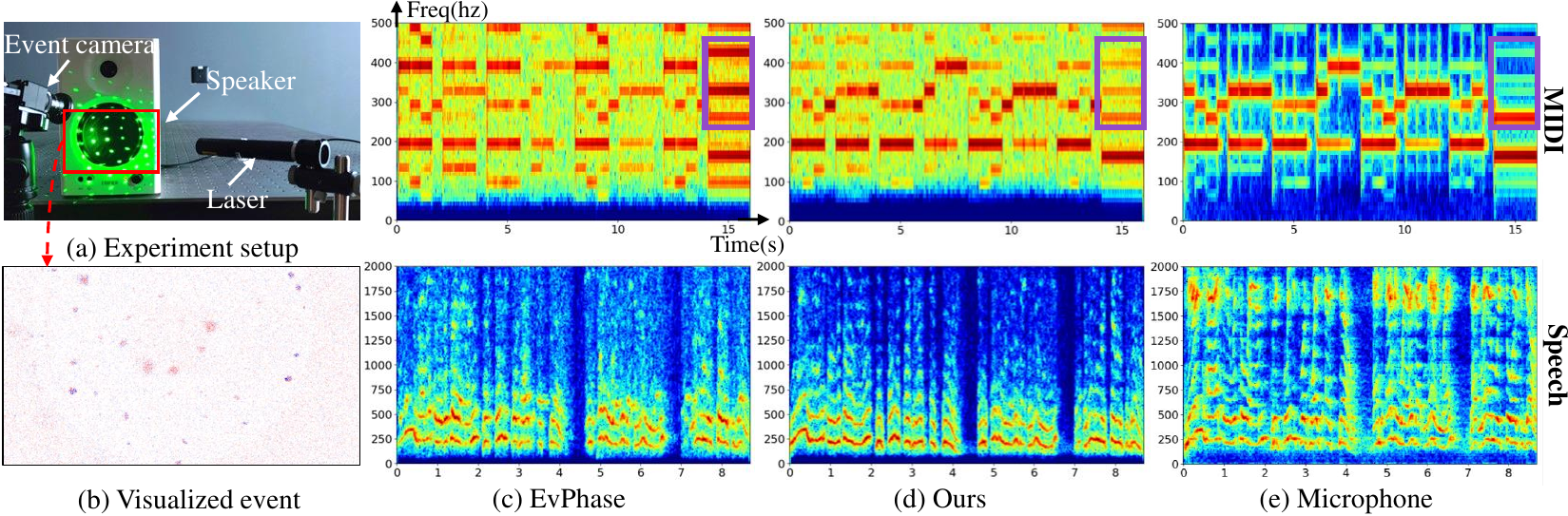}
    \caption{Qualitative comparison results on the real-world data of a speaker. \textbf{Audio is provided in the supplementary.}}
    \label{fig_speaker}
\end{figure*}

\label{sec:synthetic Experiments}
Synthetic data was generated using audio samples from the TIMIT dataset~\cite{fisher1986darpa}. High-frame-rate videos at $10,000$ FPS with a resolution of $640\times 360$ were created, and events were simulated using V2E. For the RGB method, videos are downsampled to $2,500$ FPS to replicate real-world conditions with limited bandwidth. Evaluation metrics included Signal-to-Noise Ratio (SNR) and Speech Intelligibility (STOI)~\cite{STOI}. To note that, even though our model is trained with an input sampling rate of 4kHz, as discussed in \cite{zubic2024state}, the mamba-based structure demonstrates strong adaptability when the frequency of the input data changes, and our method can easily extended to 8kHz by increasing the sampling rate at testing time. 

\cref{tab1} presents quantitative results for the synthetic sequences. The SNR and STOI metrics demonstrate that the proposed method outperforms other approaches on most sequences. Compared to EvPhase~\cite{dorn2018efficient}, our approach achieves an improvement of 1.293 dB in SNR and 0.105 in STOI. Furthermore, since the filter parameters in EvPhase~\cite{dorn2018efficient} are fixed and not adaptable to the randomly generated vibration directions, it performs suboptimally on the 'Female-SA1' sequence. In contrast, our method does not require manual adjustments for the direction of movement. Due to the limited sampling rate, RGBPhase~\cite{davis2014visual} produces recovered signals that are inferior to event-based methods in terms of frequency bandwidth and signal coherence. 



\subsubsection{Results on the real-world data}
To make the comparison in real-world scenarios, we captured several sequences with the Prophesee EVK4  camera (resolution of $1280\times 720$), paired with a $4$mm lens and a $5$mW laser matrix. We have evaluated four different setups, shown in \cref{fig_chipbag,fig_speaker,fig_diff_direction,fig_large_fov}. 
Additionally, as reference signals, audios from the scene were recorded using a microphone.

\cref{fig_chipbag} and \cref{fig_speaker} shows spectrograms of the recovered sound. Similar to \cite{davis2014visual}, we conducted experiments on a chipbag hit by sound waves, as shown in \cref{fig_chipbag}. 
\cref{fig_chipbag} (b) shows the accumulated events within a 250ms temporal window. The top row of \cref{fig_chipbag} (c), (d), and (e) presents the recovery results for the MIDI audio alongside the corresponding microphone recordings. Musical instruments and vocal performances typically consist of a fundamental frequency and harmonics that are integer multiples of that fundamental. The audio recovered by our method closely matches the microphone recordings regarding the fundamental frequency, whereas the harmonics in the results obtained by EvPhase~\cite{dorn2018efficient} magnify the high-frequency component, highlighted by the purple boxes.

\cref{fig_speaker} shows the result of sound directly output from a speaker. In comparison to the chipbag, it generates a larger vibration amplitude, resulting in clear audio signals. The audio spectrogram recovered by our method more closely resembles the results obtained from the microphone recordings, whereas the EvPhase~\cite{dorn2018efficient} method exhibits similar issues to those observed in the chipbag results (\cref{fig_chipbag}). 

\subsection{Discussion}

\begin{figure*}[htbp]
    \centering
    \includegraphics[width=1\textwidth]{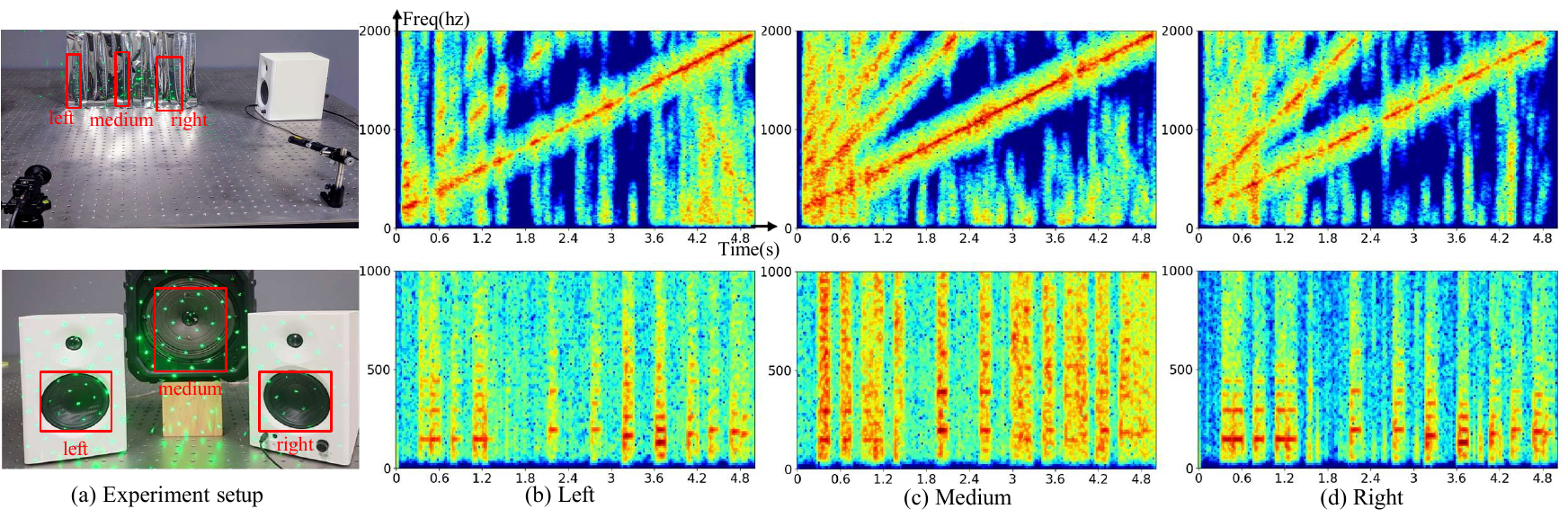}
    \caption{Capture objects from a distance to obtain a large field of view. \textbf{Top:} Capture glitter papers while playing chirp audio. \textbf{Bottom:} Capture multiple speakers to recover stereo audio. The left and right speakers play left and right channels respectively, while the medium speaker plays a mixed mono channel.  \textbf{Audio is provided in the supplementary.}}
    \label{fig_large_fov}
\end{figure*}

\textbf{Large field of view.} To demonstrate our system can cover a large field of view, we collected multiple sets of data from a long distance. As shown on the top of~\cref{fig_large_fov}, we captured glitter papers while playing chirp audio with a frequency range of 200 to 2KHz. \cref{fig_large_fov} (b), (c), and (d) illustrate the audios recovered by our model, demonstrating that signals can be recovered from various positions within the scene.

We further attempted to separate different sound sources in the scene. As shown at the bottom of~\cref{fig_large_fov}, stereo 8-bit music was played through the left and right speakers, while the central speaker played a mixed mono channel. We successfully recovered the individual audio signals from the side speakers and reconstructed a stereo audio output.

\begin{table}[!t]
    \centering
    \caption{Compare the computational load and memory usage of sparse convolution-based ResNet18 and traditional ResNet18 on EvMic.}
   \begin{tabular}{lccc}
   \toprule
    Model                 & ~ & VRAM (G)        & Flops (G)        \\ \hline
    w/o Sparse Convolution & ~ & 22.715         & 1171.226        \\ 
    w Sparse Convolution  & ~ & \textbf{7.461} & \textbf{51.229} \\ \bottomrule
\end{tabular}
    \label{tab2}
\end{table}

\begin{table}[!t]
    \centering
    \caption{Ablation analysis for our modules on the synthetic data.}
    \begin{tabular}{l@{\hskip 20pt} @{\hskip 20pt}c@{\hskip 20pt}c}
    \toprule
        Model & SNR↑ & STOI↑ \\ \hline
        SPconv + Transformer & -0.195 & 0.437 \\
        SPconv + LSTM & 0.015 & 0.453 \\ 
        SPconv + Mamba & 0.309 & 0.474 \\ \hline
        SPconv + SAB + Mamba & \textbf{1.214} & \textbf{0.481} \\ \bottomrule
    \end{tabular}
    \label{tab3}
\end{table}
%
\noindent\textbf{Sparse convolution.} To demonstrate the effectiveness of sparse convolution, we calculate the average computational load and memory usage at inference time on the EvMic dataset. We use input data with the dimensions \(x\in  R^{C\times T\times H\times W} \), where \(C=2\) is the polarity of the events, \(H=256\) and \(W=344\) represent the spatial dimensions, and \(T=2k\) denotes the temporal length. 

\begin{figure*}[htbp]
    \centering
    \includegraphics[width=1\textwidth]{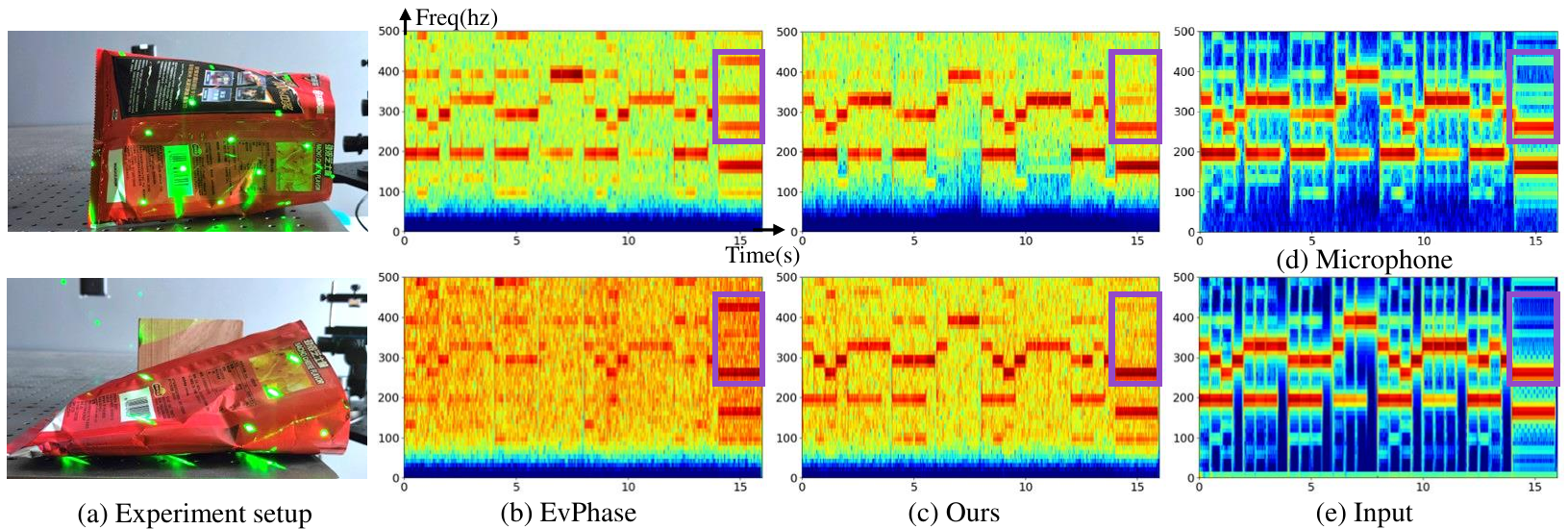}
    \caption{Ablation analysis for different vibration direction. The object is placed in different orientations to produce various vibration directions. \textbf{Audio is provided in the supplementary.}}
    \label{fig_diff_direction}
\end{figure*}

\cref{tab2} illustrates the memory usage and computational load of the two methods. The sparse backbone utilizes the sparsity of event data for lightweight visual feature extraction, using only 32.65\% of the memory and 4.37\% of the computational load compared to traditional CNNs.\\
\textbf{Temporal modeling and spatial aggregation.} We conduct the ablation analysis on the remaining modules. We quantitatively compare different temporal modeling approaches~\cite{vaswani2017attention, gu2023mamba,hochreiter1997long}, and the inclusion of the spatial aggregation block (SAB) in~\cref{tab3}. The transformer performs poorly in metrics. Qualitative results are available in our supplementary materials. Mamba outperforms LSTM by achieving a 0.294dB improvement in SNR and a 0.021 increase in STOI, demonstrating its exceptional performance in modeling long-term temporal information. We further validated the effectiveness of the SAB module. The inclusion of the SAB resulted in a 0.905dB improvement in SNR and a 0.007 increase in STOI. \\
\textbf{Adaptability to different vibration directions.} In the experiments shown in~\cref{fig_diff_direction}(a), we altered the vibration direction by changing the orientation of the object's surface. We keep the parameters of the EvPhase~\cite{dorn2018efficient} method unchanged (\cref{sec:Comparation with other methods}). \cref{fig_diff_direction}(b) and (c) illustrate the signals recovered by both methods for two different vibration directions. Compared to the signals recovered in the original direction, our method incurs a slight quality drop along the new vibration direction. In contrast, the EvPhase~\cite{dorn2018efficient} method recovers a poor signal in the altered vibration direction due to the single spatial orientation of its filter.

\section{Conclusion}
In this work, we explored a learning-based pipeline for event-based non-contact sound recovery for the first time. We used Blender to synthesize a dataset named EvMic and designed a model specifically for this task. Our approach enhanced the quality of the recovered signals by effectively modeling spatial-temporal information. However, some limitations still need to be addressed. First, there is a gap between the existing event simulator and the events captured in reality, which leads to a decline in performance on real data. Besides, the generation of events is influenced by the gradient. Therefore, the power of the laser and the lighting conditions will affect the number of captured events. In future work, we aim to refine our acquisition system and explore the incorporation of prior knowledge from generative models to further enhance the quality of the signals.

\clearpage
{
    \small
    \bibliographystyle{ieeenat_fullname}
    \bibliography{main}
}

\clearpage
\setcounter{page}{1}
\maketitlesupplementary
\appendix
\renewcommand\thefigure{A\arabic{figure}}
\renewcommand\thetable{A\arabic{table}}  
\renewcommand\theequation{A\arabic{equation}}
\setcounter{section}{0}
\setcounter{equation}{0}
\setcounter{table}{0}
\setcounter{figure}{0}

\section{Supplementary audio results and video}
To facilitate a comprehensive comparison among different methods, our supplementary material includes the audio results of our proposed method, EvPhase~\cite{dorn2018efficient}, and RGBPhase~\cite{davis2014visual} (available in the "audio results" folder). Additionally, we provide a video that demonstrates our main pipeline along with our real-world experiments.

\begin{figure}[htbp]
    \centering
    \includegraphics[width=0.49\textwidth]{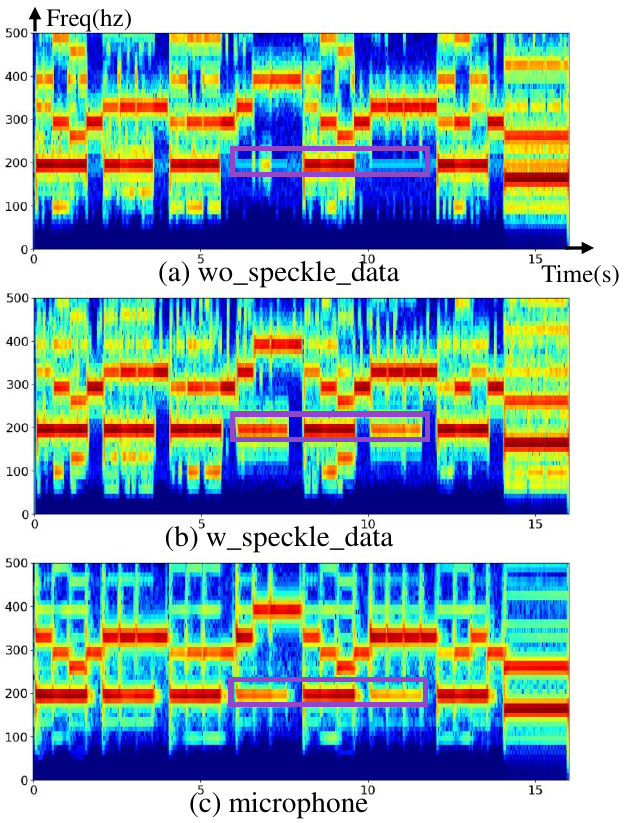}
    \caption{Qualitative comparison for models trained w or w/o speckle data.}
    \label{speckle_qualitative}
\end{figure}

\section{Synthetic speckle data.} In real-world scenarios, we employ a laser matrix to enhance surface gradients, with events primarily occurring at the laser speckles. However, the absence of speckle patterns in our training dataset adversely affects the model's generalization. To bridge the gap between simulated and real data, we captured real scene recordings and extracted the speckle textures from the accumulated event frames, as illustrated in Fig.~\ref{fig2}(b).

Specifically, we extracted multiple patches from the accumulated event frames, each centered on a speckle with a resolution of $32\times 32$.  These patches were then upsampled to a resolution of $128\times 128$ using bilinear interpolation. Then we initialized a black background with a resolution of $2560\times 1440$, using the audio to control the inter-frame displacements, allowing the patches to vibrate in random directions. To ensure diversity in the vibration amplitude, we multiplied the audio by a random gain ranging from 0.5 to 2, and the resulting images were downsampled to a resolution of $640\times 360$. After generating the video sequences, we employed V2E~\cite{hu2021v2e} to simulate the events. The resulting speckle sequences are utilized to finetune the spatial aggregation block. 

Fig.~\ref{speckle_qualitative} illustrates the recovered audio using models trained with and without the synthetic speckle data. The model trained with speckle data achieved a more comprehensive signal recovery, as highlighted in the purple box.

\begin{figure}[htbp]
    \centering
    \includegraphics[width=0.49\textwidth]{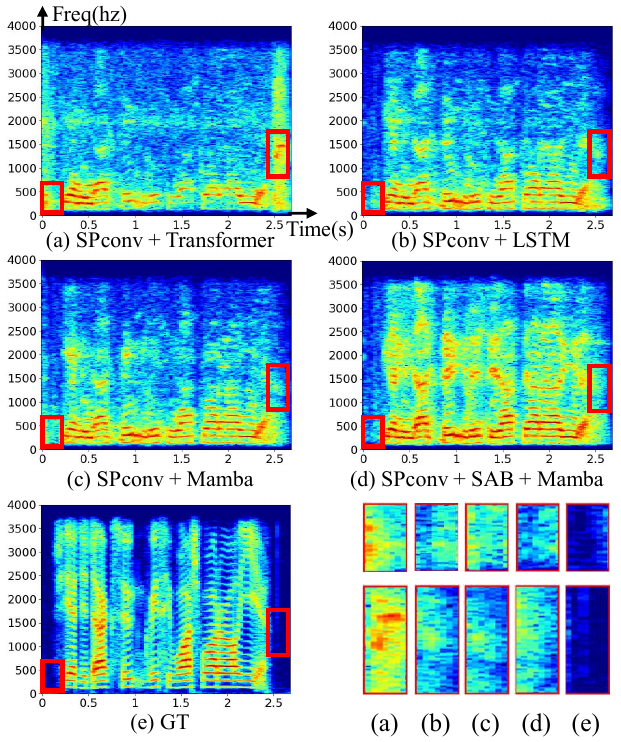}
    \caption{Qualitative results for ablation analysis.}
    \label{ablation_qualitative}
\end{figure}

\begin{figure*}[htbp]
    \centering
    \includegraphics[width=1\textwidth]{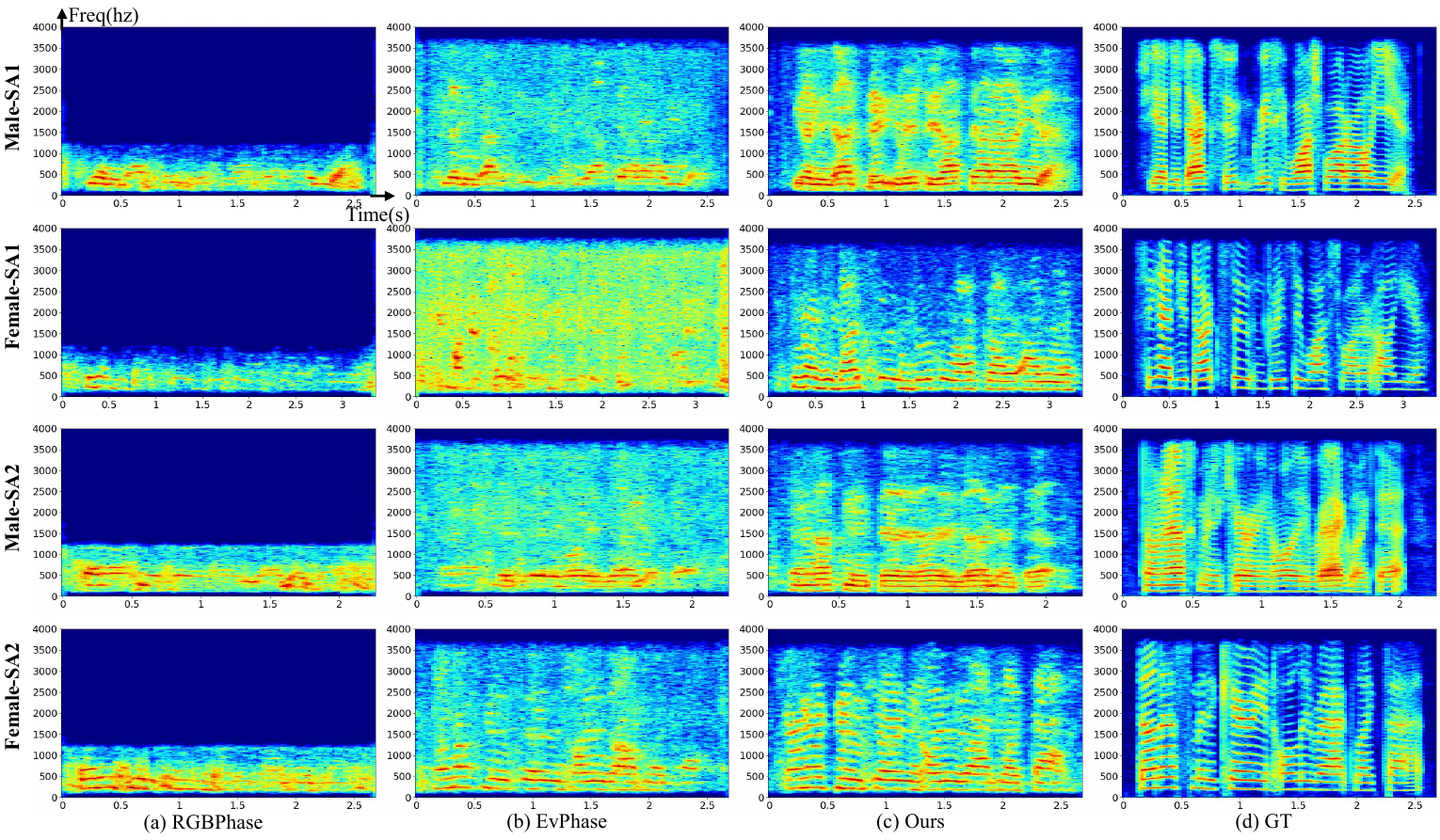}
    \caption{Qualitative comparison results of our model with other methods on the synthetic data.}
    \label{syn_qualitative}
\end{figure*}

\begin{figure*}[htbp]
    \centering
    \includegraphics[width=1\textwidth]{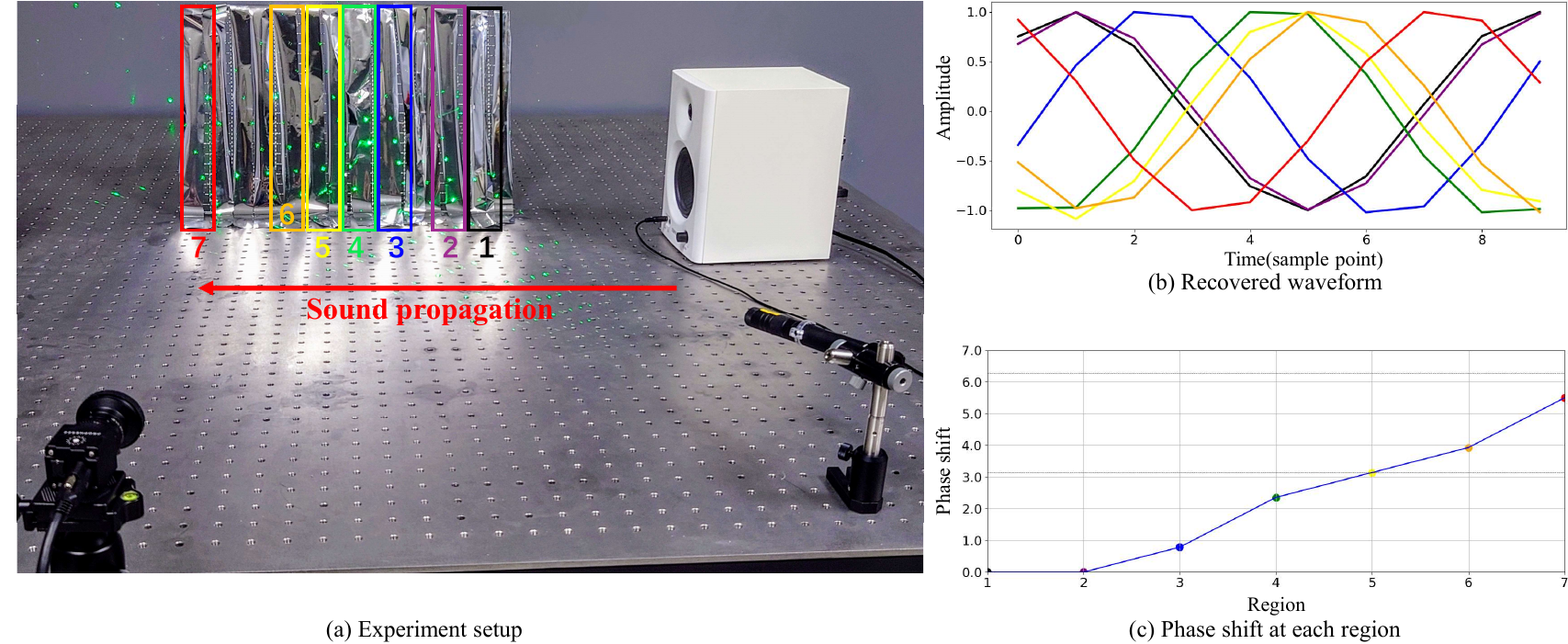}
    \caption{Phase shift introduced by sound propagation.}
    \label{phase_shift}
\end{figure*}

\section{Additional ablation analysis}
As mentioned in the main paper, the transformer performs poorly in metrics among the compared approaches. The low correlation between the noisy static frames results in evenly distributed attention weights. This leads to their influence from other frames within the time window. As shown in Fig~\ref{ablation_qualitative}, the transformer incorrectly restored sound in the absence of any audio (highlighted by the red boxes). \\

\section{Qualitative results on synthetic data}
Fig.~\ref{syn_qualitative} illustrates the qualitative results on synthetic data, presented in spectrograms. RGBPhase~\cite{davis2014visual} is limited by the sampling rate and can only recover information below 1 kHz. Due to the single spatial orientation of the filter, EvPhase~\cite{dorn2018efficient} performs poorly in "Female-SA1" with fixed parameters. In contrast, our method achieves a higher signal-to-noise ratio and better integrity in the recovered signals compared to EvPhase~\cite{dorn2018efficient}.

\section{Quantity results on the real-world data.}
We use audios recorded by a microphone as ground truth to calculate quantitative metrics on real data, as shown in Tab. \ref{real_world_metrics}. However, several factors affect this approach: (1) room reverberation; (2) temporal misalignment between the microphone recordings and the recovered audios; and (3) varying object responses to different frequencies, leading to discrepancies between their vibrations and the microphone recordings. Consequently, the microphone recordings do not represent ideal ground truth.

\begin{table*}[!ht]
    \centering
    \caption{Quantity results on the real-world data. }
\begin{tabular}{lllllcclcclcclcccc}
\toprule
 & \multicolumn{4}{l}{\multirow{2}{*}{Methods}} & \multicolumn{2}{c}{Chipbag-Speech} &  & \multicolumn{2}{c}{Chipbag-MIDI} &  & \multicolumn{2}{c}{Speaker-Speech} &  & \multicolumn{2}{c}{Speaker-MIDI} & \multicolumn{2}{c}{Average} \\ \cline{6-18} 
 & \multicolumn{4}{l}{} & SNR↑ & STOI↑ &  & SNR↑ & STOI↑ &  & SNR↑ & STOI↑ &  & SNR↑ & STOI↑ & SNR↑ & STOI↑ \\ \hline
 & \multicolumn{4}{l}{EvPhase} & -1.540 & 0.290 &  & 1.122 & - &  & -2.963 & 0.501 &  & 0.681 & - & -1.350 & 0.396 \\ 
 & \multicolumn{4}{l}{Ours(8kHz)} & \textbf{-0.511} & \textbf{0.383} &  & \textbf{3.867} & - &  & \textbf{-2.866} & \textbf{0.506} &  & \textbf{3.338} & - & \textbf{0.957} & \textbf{0.445} \\ \bottomrule
\end{tabular}

\label{real_world_metrics}
\end{table*}

\section{Inference times for all models.}
We measured the inference time for all compared methods on the synthetic data, as shown in the 
 Tab. \ref{time}. To ensure a fair comparison, all experiments were conducted on an Intel 14700K CPU. 
Here, ``ours-$n$" indicates the use of $n$ speckles for inference. When utilizing a single speckle, our inference speed is on par with EvPhase. The comparisons in our main paper are conducted with eight speckles. Under this setting, our method is slower than EvPhase but still offers a significant speed advantage over RGBPhase.

\begin{table}[!ht]
\caption{Inference time for all methods.}
\begin{tabular}{lllll}
\toprule
Methods      & RGBPhase     & EvPhase & Ours-1 & Ours-8 \\ \hline
time(s) & 744.24 & 29.54   & 30.13  & 132.36 \\ \bottomrule
\label{time}
\end{tabular}
\end{table}

\section{Analyze phase shift in sound propagation.}
Our imaging system can achieve a large field of view, allowing us to analyze the phase shift caused by sound propagation in space. We captured a row of glitter papers and attempted to analyze this phase shift. The experimental setup is shown in Fig.~\ref{phase_shift} (a).  We played a 1,000 Hz sine wave through the speaker on the right, allowing the sound to propagate toward the left. Using our method, we reconstructed the audio at a sampling rate of 8000 Hz. The reconstructed waveforms for different regions (marked with different colors) are shown in Fig.~\ref{phase_shift} (b). We plotted the phase shift variation during the sound propagation process as a line graph, as illustrated in Fig.~\ref{phase_shift} (c).

\end{document}